\documentclass[aps,prl,preprint,superscriptaddress,showpacs,preprintnumbers,amsmath,amssymb]{revtex4-1}

\usepackage{xspace}
\usepackage{graphicx} % Include figure files
\usepackage{dcolumn}  % Align table columns on decimal point
\usepackage[T1]{fontenc} % if needed

\graphicspath{{ps}}

\newcommand{\lcp}{\Lambda_c^+}

\newcommand{\lc}{\Lambda_c}
\newcommand{\br}{{\mathcal B}}
\newcommand{\fb}{fb$^{-1}$}

\newcommand{\dppilc}{D^{(\ast)-}\overline{p}\pi^+\lcp}
\newcommand{\dppi}{D^{(\ast)-}\overline{p}\pi^+}
\newcommand{\dppinc}{D^{(\ast)}p\pi}

\newcommand{\fbias}{f_{\rm bias}}

\newcommand{\mmiss}{M_{\textrm{miss}}}

\newcommand{\lcpkpi}{\Lambda_c^+\to pK^-\pi^+}
\newcommand{\pkpi}{pK^-\pi^+}

\newcommand{\gev}{\ensuremath{\mathrm{\,Ge\kern -0.1em V}}\xspace}
\newcommand{\mev}{\ensuremath{\mathrm{\,Me\kern -0.1em V}}\xspace}
\newcommand{\gevc}{\ensuremath{{\mathrm{\,Ge\kern -0.1em V\!/}c}}\xspace}
\newcommand{\mevc}{\ensuremath{{\mathrm{\,Me\kern -0.1em V\!/}c}}\xspace}
\newcommand{\gevcc}{\ensuremath{{\mathrm{\,Ge\kern -0.1em V\!/}c^2}}\xspace}
\newcommand{\gevsqcq}{\ensuremath{{\mathrm{\,Ge\kern -0.1em V^2\!/}c^4}}\xspace}
\newcommand{\mevcc}{\ensuremath{{\mathrm{\,Me\kern -0.1em V\!/}c^2}}\xspace}

\begin{document}

\vspace*{-3\baselineskip}
\resizebox{!}{3cm}{\includegraphics{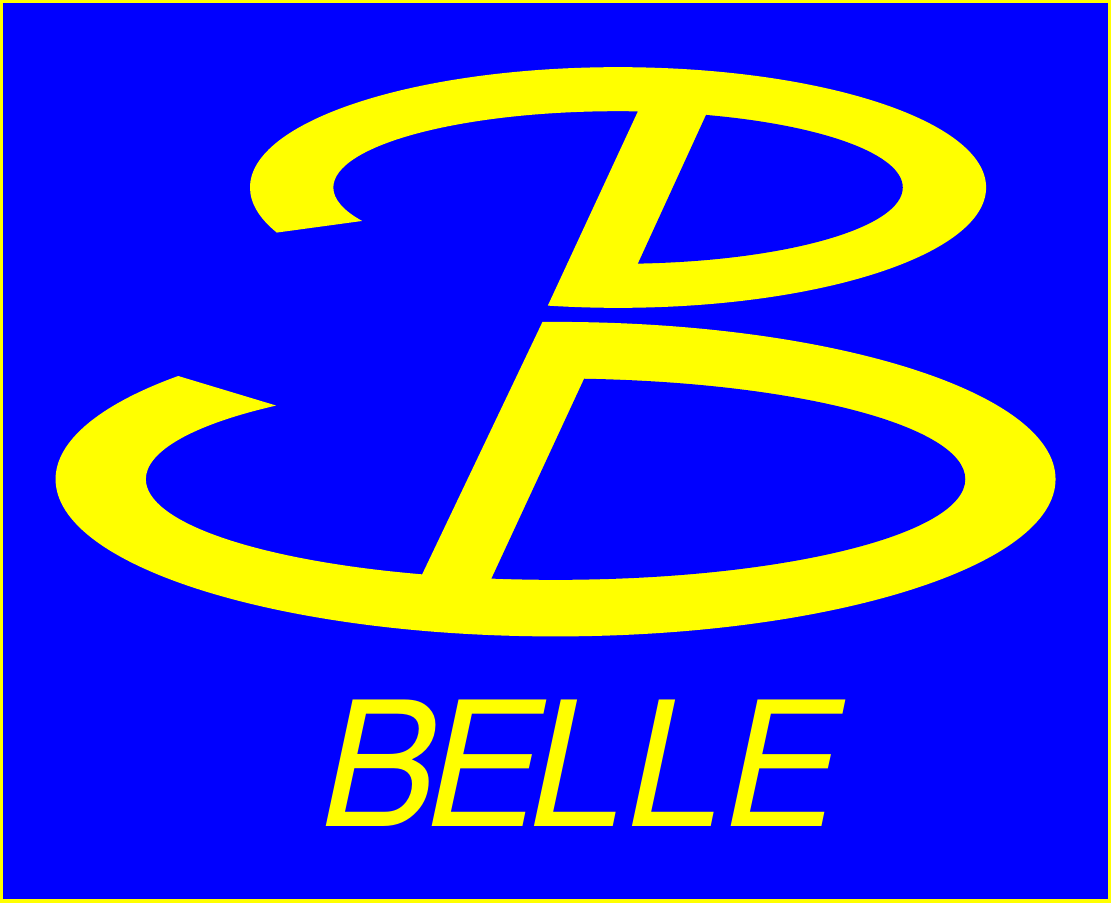}}

\preprint{\vbox{ \hbox{Belle Preprint 2013-30}
    \hbox{KEK Preprint 2013-58}
}}

\title{\boldmath \quad\\[1.0cm] Measurement of the Branching Fraction $\br(\lcpkpi)$}

%%% Paper:    Lambda_c+ -> p K- pi+
%%% Journal:  Physical Review Letters
%%% Contacts: A. Zupanc (zupanc@ekp.uni-karlsruhe.de)
%%%           C. Bartel (cbartel@ekp.uni-karlsruhe.de)
%%%           N. Gabyshev (gabyshev@inp.nsk.su)
%%% Non-responding authors or those who said NO are commented out.
%%% ====================================================================
%%% Click the RELOAD button on your web browser to see the updated file.
%%% ====================================================================
%%% Use \input{author} to insert this material into your latex file.
%%%%% Force institutions to appear in alphabetical order when typeset.
\noaffiliation
\affiliation{University of the Basque Country UPV/EHU, 48080 Bilbao}
\affiliation{Beihang University, Beijing 100191}
\affiliation{University of Bonn, 53115 Bonn}
\affiliation{Budker Institute of Nuclear Physics SB RAS and Novosibirsk State University, Novosibirsk 630090}
\affiliation{Faculty of Mathematics and Physics, Charles University, 121 16 Prague}
%%%\affiliation{Chiba University, Chiba 263-8522}
\affiliation{University of Cincinnati, Cincinnati, Ohio 45221}
\affiliation{Deutsches Elektronen--Synchrotron, 22607 Hamburg}
\affiliation{Department of Physics, Fu Jen Catholic University, Taipei 24205}
\affiliation{Justus-Liebig-Universit\"at Gie\ss{}en, 35392 Gie\ss{}en}
%%%\affiliation{Gifu University, Gifu 501-1193}
%%%\affiliation{II. Physikalisches Institut, Georg-August-Universit\"at G\"ottingen, 37073 G\"ottingen}
%%%\affiliation{The Graduate University for Advanced Studies, Hayama 240-0193}
\affiliation{Gyeongsang National University, Chinju 660-701}
\affiliation{Hanyang University, Seoul 133-791}
\affiliation{University of Hawaii, Honolulu, Hawaii 96822}
\affiliation{High Energy Accelerator Research Organization (KEK), Tsukuba 305-0801}
%%%\affiliation{Hiroshima Institute of Technology, Hiroshima 731-5193}
\affiliation{IKERBASQUE, Basque Foundation for Science, 48011 Bilbao}
%%%\affiliation{University of Illinois at Urbana-Champaign, Urbana, Illinois 61801}
\affiliation{Indian Institute of Technology Guwahati, Assam 781039}
\affiliation{Indian Institute of Technology Madras, Chennai 600036}
%%%\affiliation{Indiana University, Bloomington, Indiana 47408}
\affiliation{Institute of High Energy Physics, Chinese Academy of Sciences, Beijing 100049}
\affiliation{Institute of High Energy Physics, Vienna 1050}
\affiliation{Institute for High Energy Physics, Protvino 142281}
%%%\affiliation{Institute of Mathematical Sciences, Chennai 600113}
\affiliation{INFN - Sezione di Torino, 10125 Torino}
\affiliation{Institute for Theoretical and Experimental Physics, Moscow 117218}
\affiliation{J. Stefan Institute, 1000 Ljubljana}
\affiliation{Kanagawa University, Yokohama 221-8686}
\affiliation{Institut f\"ur Experimentelle Kernphysik, Karlsruher Institut f\"ur Technologie, 76131 Karlsruhe}
\affiliation{Kavli Institute for the Physics and Mathematics of the Universe (WPI), University of Tokyo, Kashiwa 277-8583}
\affiliation{Korea Institute of Science and Technology Information, Daejeon 305-806}
\affiliation{Korea University, Seoul 136-713}
\affiliation{Kyoto University, Kyoto 606-8502}
\affiliation{Kyungpook National University, Daegu 702-701}
\affiliation{\'Ecole Polytechnique F\'ed\'erale de Lausanne (EPFL), Lausanne 1015}
\affiliation{Faculty of Mathematics and Physics, University of Ljubljana, 1000 Ljubljana}
\affiliation{Luther College, Decorah, Iowa 52101}
\affiliation{University of Maribor, 2000 Maribor}
\affiliation{Max-Planck-Institut f\"ur Physik, 80805 M\"unchen}
\affiliation{School of Physics, University of Melbourne, Victoria 3010}
\affiliation{Moscow Physical Engineering Institute, Moscow 115409}
%%%\affiliation{Moscow Institute of Physics and Technology, Moscow Region 141700}
\affiliation{Graduate School of Science, Nagoya University, Nagoya 464-8602}
\affiliation{Kobayashi-Maskawa Institute, Nagoya University, Nagoya 464-8602}
%%%\affiliation{Nara University of Education, Nara 630-8528}
\affiliation{Nara Women's University, Nara 630-8506}
%%%\affiliation{National Central University, Chung-li 32054}
\affiliation{National United University, Miao Li 36003}
\affiliation{Department of Physics, National Taiwan University, Taipei 10617}
\affiliation{H. Niewodniczanski Institute of Nuclear Physics, Krakow 31-342}
\affiliation{Nippon Dental University, Niigata 951-8580}
\affiliation{Niigata University, Niigata 950-2181}
\affiliation{University of Nova Gorica, 5000 Nova Gorica}
\affiliation{Osaka City University, Osaka 558-8585}
%%%\affiliation{Osaka University, Osaka 565-0871}
\affiliation{Pacific Northwest National Laboratory, Richland, Washington 99352}
\affiliation{Panjab University, Chandigarh 160014}
\affiliation{Peking University, Beijing 100871}
\affiliation{University of Pittsburgh, Pittsburgh, Pennsylvania 15260}
%%%\affiliation{Punjab Agricultural University, Ludhiana 141004}
%%%\affiliation{Research Center for Electron Photon Science, Tohoku University, Sendai 980-8578}
%%%\affiliation{Research Center for Nuclear Physics, Osaka University, Osaka 567-0047}
%%%\affiliation{RIKEN BNL Research Center, Upton, New York 11973}
%%%\affiliation{Saga University, Saga 840-8502}
\affiliation{University of Science and Technology of China, Hefei 230026}
\affiliation{Seoul National University, Seoul 151-742}
%%%\affiliation{Shinshu University, Nagano 390-8621}
\affiliation{Soongsil University, Seoul 156-743}
\affiliation{Sungkyunkwan University, Suwon 440-746}
\affiliation{School of Physics, University of Sydney, NSW 2006}
\affiliation{Tata Institute of Fundamental Research, Mumbai 400005}
\affiliation{Excellence Cluster Universe, Technische Universit\"at M\"unchen, 85748 Garching}
\affiliation{Toho University, Funabashi 274-8510}
\affiliation{Tohoku Gakuin University, Tagajo 985-8537}
\affiliation{Tohoku University, Sendai 980-8578}
\affiliation{Department of Physics, University of Tokyo, Tokyo 113-0033}
\affiliation{Tokyo Institute of Technology, Tokyo 152-8550}
\affiliation{Tokyo Metropolitan University, Tokyo 192-0397}
\affiliation{Tokyo University of Agriculture and Technology, Tokyo 184-8588}
\affiliation{University of Torino, 10124 Torino}
%%%\affiliation{Toyama National College of Maritime Technology, Toyama 933-0293}
\affiliation{CNP, Virginia Polytechnic Institute and State University, Blacksburg, Virginia 24061}
\affiliation{Wayne State University, Detroit, Michigan 48202}
\affiliation{Yamagata University, Yamagata 990-8560}
\affiliation{Yonsei University, Seoul 120-749}
  \author{A.~Zupanc}\affiliation{J. Stefan Institute, 1000 Ljubljana} % Ljubljana
  \author{C.~Bartel}\affiliation{Institut f\"ur Experimentelle Kernphysik, Karlsruher Institut f\"ur Technologie, 76131 Karlsruhe} % Karlsruhe
  \author{N.~Gabyshev}\affiliation{Budker Institute of Nuclear Physics SB RAS and Novosibirsk State University, Novosibirsk 630090} % BINP
  \author{I.~Adachi}\affiliation{High Energy Accelerator Research Organization (KEK), Tsukuba 305-0801} % KEK
% \author{K.~Adamczyk}\affiliation{H. Niewodniczanski Institute of Nuclear Physics, Krakow 31-342} % Krakow
  \author{H.~Aihara}\affiliation{Department of Physics, University of Tokyo, Tokyo 113-0033} % Tokyo
% \author{K.~Arinstein}\affiliation{Budker Institute of Nuclear Physics SB RAS and Novosibirsk State University, Novosibirsk 630090} % BINP
% \author{Y.~Arita}\affiliation{Graduate School of Science, Nagoya University, Nagoya 464-8602} % Nagoya
  \author{D.~M.~Asner}\affiliation{Pacific Northwest National Laboratory, Richland, Washington 99352} % PNNL
% \author{T.~Aso}\affiliation{Toyama National College of Maritime Technology, Toyama 933-0293} % Toyama
  \author{V.~Aulchenko}\affiliation{Budker Institute of Nuclear Physics SB RAS and Novosibirsk State University, Novosibirsk 630090} % BINP
  \author{T.~Aushev}\affiliation{Institute for Theoretical and Experimental Physics, Moscow 117218} % ITEP
% \author{T.~Aziz}\affiliation{Tata Institute of Fundamental Research, Mumbai 400005} % Tata
  \author{A.~M.~Bakich}\affiliation{School of Physics, University of Sydney, NSW 2006} % Sydney
  \author{A.~Bala}\affiliation{Panjab University, Chandigarh 160014} % Panjab
% \author{Y.~Ban}\affiliation{Peking University, Beijing 100871} % Peking
% \author{E.~Barberio}\affiliation{School of Physics, University of Melbourne, Victoria 3010} % Melbourne
% \author{M.~Barrett}\affiliation{University of Hawaii, Honolulu, Hawaii 96822} % Hawaii
% \author{W.~Bartel}\affiliation{Deutsches Elektronen--Synchrotron, 22607 Hamburg} % DESY
% \author{A.~Bay}\affiliation{\'Ecole Polytechnique F\'ed\'erale de Lausanne (EPFL), Lausanne 1015} % Lausanne
% \author{I.~Bedny}\affiliation{Budker Institute of Nuclear Physics SB RAS and Novosibirsk State University, Novosibirsk 630090} % BINP
% \author{P.~Behera}\affiliation{Indian Institute of Technology Madras, Chennai 600036} % IITM
% \author{M.~Belhorn}\affiliation{University of Cincinnati, Cincinnati, Ohio 45221} % Cincinnati
% ...
% BINP
% \author{T.~Zivko}\affiliation{J. Stefan Institute, 1000 Ljubljana} % Ljubljana
% \author{N.~Zwahlen}\affiliation{\'Ecole Polytechnique F\'ed\'erale de Lausanne (EPFL), Lausanne 1015} % Lausanne
% \author{O.~Zyukova}\affiliation{Budker Institute of Nuclear Physics SB RAS and Novosibirsk State University, Novosibirsk 630090} % BINP
\collaboration{The Belle Collaboration}

\begin{abstract}
We present the first model-independent measurement of the absolute branching fraction of the 
$\lcpkpi$ decay using a data sample of 978~\fb\ collected with the Belle detector at the KEKB 
asymmetric-energy $e^+e^-$ collider. The number of $\lcp$ baryons is determined by reconstructing 
the recoiling $\dppi$ system in events of the type $e^+e^-\to \dppi\lcp$. The branching fraction
is measured to be $\br(\lcpkpi)=(6.84\pm0.24{}^{+0.21}_{-0.27})\%$, where the first and second 
uncertainties are statistical and systematic, respectively. 
\end{abstract}


\maketitle

{\renewcommand{\thefootnote}{\fnsymbol{footnote}}}
\setcounter{footnote}{0}

The hadronic decay $\lcpkpi$ is the reference mode for the measurements of branching fractions of the $\lcp$ baryon to any other 
final state~\cite{Beringer:1900zz}. In addition, this is the most common decay mode in studies where a $\lcp$ baryon is 
included in the final state of the decay chain, such as the exclusive and inclusive decay rate measurements of $b$-flavored mesons 
and baryons or the measurements of fragmentation fractions of charm and bottom quarks. 
The Particle Data Group combines several measurements from the ARGUS and CLEO 
collaborations~\cite{Albrecht:1988an,Crawford:1991at,Albrecht:1992he,Albrecht:1991bu,Bergfeld:1994gt,Albrecht:1996gr,Avery:1990bc} 
to determine $\br(\lcpkpi)=(5.0\pm1.3)\%$, where the dominant contribution to the quoted uncertainty originates 
from the model dependence of the branching fraction extraction~\footnote{For a review of the model-dependent 
measurements, see Sec. ``$\lcp$ branching fractions'' on page 1386 in Ref.~\cite{Beringer:1900zz}.}. A precise measurement of the branching fraction $\br(\lcpkpi)$ can therefore significantly improve the precision of branching fractions of other $\lcp$ decays and also those of decays 
of $b$-flavored mesons and baryons involving $\lcp$.

In this Letter, we present the first model-independent measurement of the absolute branching fraction of 
$\lcpkpi$ decays~\footnote{Throughout this paper, charge-conjugate modes are included.} 
that improves the precision of previous model-dependent measurements by a 
factor of five. We use a data sample, corresponding to an integrated luminosity of 978~\fb, collected at or near the $\Upsilon(nS)$ ($n=1,2,3,4,5$) 
resonances with the Belle detector at the KEKB asymmetric-energy $e^+e^-$ collider.

The absolute branching fraction of the $\lcpkpi$ decay is given by 
\begin{equation}
 \br(\lcpkpi) = \frac{N(\lcpkpi)}{N_{\textrm{inc}}^{\lc}\fbias\varepsilon(\lcpkpi)},
 \label{eq:abs_br}
\end{equation}
where $N_{\textrm{inc}}^{\lc}$ is the number of inclusively reconstructed $\lcp$ baryons,  
$N(\lcpkpi)$ is the number of reconstructed $\lcpkpi$ decays within the inclusive $\lcp$ sample, 
$\varepsilon(\lcpkpi)$ is the reconstruction efficiency of $\lcpkpi$ decays within the inclusive $\lcp$ sample,
and the factor $\fbias$ takes into account potential dependence of the inclusive $\lcp$ reconstruction 
efficiency on the $\lcp$ decay mode.

The $e^+e^- \to c\bar{c}$ events that contain $\lcp$ baryons produced through the reactions 
$e^+e^-\to c\bar{c}\to \dppilc$ are fully reconstructed in two steps. In the first, 
no requirements are placed on the daughters of the $\lcp$ baryons in order to obtain an inclusive sample 
of $\lcp$ events that is used as the denominator in the calculation of the branching fraction. The number of inclusively 
reconstructed $\lcp$ baryons is extracted from the distribution of events in the missing mass recoiling against 
the $\dppi$ system, $\mmiss(\dppinc) = \sqrt{p^2_{\textrm{miss}}(\dppinc)}$, where 
$p_{\textrm{miss}}(\dppinc)  =  p_{e^+} + p_{e^-} - p_{D^{(\ast)}} - p_{p} - p_{\pi}$ is the missing four-momentum in the event.
Here, $p_{e^+}$ and $p_{e^-}$ are the known four-momenta of the colliding positron and electron beams, respectively, and $p_{D^{(\ast)}}$, 
$p_{p}$, and $p_{\pi}$ are the measured four-momenta of the reconstructed $D^{(\ast)}$, the antiproton, and the pion, 
respectively. Correctly reconstructed events produce a peak in the $\mmiss(\dppinc)$ distribution at the nominal $\lcp$ mass.
In the second step, we search for the decay products of the $\lcpkpi$ decay within the inclusive $\lcp$ sample 
reconstructed in the first step. In particular, we require that there be only three charged tracks, consistent with being a kaon, 
pion and proton, in the rest of the event.

The Belle detector is a large-solid-angle magnetic spectrometer that consists of a silicon vertex detector (SVD), a 50-layer 
central drift chamber (CDC), an array of aerogel threshold Cherenkov counters (ACC), a barrel-like arrangement of time-of-flight 
scintillation counters (TOF), and an electromagnetic calorimeter (ECL) comprised of CsI(Tl) crystals located inside a superconducting 
solenoid coil that provides a 1.5~T magnetic field.  An iron flux-return located outside of the coil is instrumented to detect 
$K_L^0$ mesons and to identify muons. The detector is described in detail elsewhere~\cite{Brodzicka:2012jm,Abashian:2000cg}. 
We use Monte Carlo (MC) events generated with EVTGEN~\cite{Lange:2001uf} and JETSET~\cite{Sjostrand:1993yb} and then processed through 
the detailed detector simulation implemented in GEANT3~\cite{Brun:1987ma}. Final state radiation from charged particles
is simulated during event generation using the PHOTOS package~\cite{Barberio:1993qi}. The simulated samples for $e^+e^-$ annihilation 
to $q\overline{q}$ ($q=u$, $d$, $s$, $c$, and $b$) are equivalent to six times the integrated luminosity of the data and are used to develop
methods to separate signal events from backgrounds, identify types of background events, determine the reconstruction efficiency
and parameterize the distributions needed for the extraction of the signal decays.

Charged particles are reconstructed with the CDC and the SVD. Each is required to have an impact parameter with 
respect to the interaction point (IP) of less than 1.5~cm along the positron beam direction and less than 0.5~cm in 
the plane transverse to the positron beam direction. A likelihood ratio for a given track to be 
a kaon, pion, or proton 
is obtained by utilizing energy-loss measurements in the CDC, 
light yield measurements from the ACC, and time-of-flight information from the TOF. Photons are detected with 
the ECL and are required to have energies in the laboratory frame of at least 50 (100) \mev in the ECL barrel 
(endcaps). Neutral pion candidates are reconstructed using photon pairs with an invariant mass between 120 
and 150~\mevcc, which corresponds to $\pm$3.2$\sigma$ around the nominal $\pi^0$ mass~\cite{Beringer:1900zz}, 
where $\sigma$ represents the resolution. Neutral kaon candidates are reconstructed using 
pairs of oppositely-charged pions with an invariant mass within $\pm20$~\mevcc ($\pm5\sigma$) of the nominal $K^0$ mass.

We reconstruct the charmed pseudoscalar mesons in the following twelve decay modes: 
$D^0\to K^-\pi^+$, $K^-\pi^+\pi^0$, $K^-\pi^+\pi^+\pi^-$, $K^-\pi^+\pi^+\pi^-\pi^0$,
$K^0_S\pi^+\pi^-$ or $K^0_S\pi^+\pi^-\pi^0$; $D^+\to K^-\pi^+\pi^+$, $K^-\pi^+\pi^+\pi^0$, $K^0_S\pi^+$, 
$K^0_S\pi^+\pi^0$, $K^0_S\pi^+\pi^+\pi^-$, or $K^+K^-\pi^+$.
In order to reject background from $e^+e^-\to B\overline{B}$ events and combinatorial background, the 
$D$ momentum in the $e^+e^-$ frame is required to be greater than 2.3 to 2.5~$\gevc$, depending on the decay mode. 
To further increase the purity of the reconstructed sample of charmed pseudoscalar mesons, we combine several variables 
into a single output variable using the NeuroBayes neural network~\cite{Feindt:2006pm}:
the distance between the decay and the production vertices of the $D$  
candidate in the transverse plane, where the $D$ production vertex is defined by the intersection of its trajectory with the IP region; 
the $\chi^2$ of the vertex fit of the $D$ candidate; the cosine of the angle between the $D$ momentum and the vector joining
its decay and production vertices in the transverse plane; for two-body $D\to K\pi$ decays, the cosine of the angle between the kaon 
momentum and the boost direction of the laboratory frame in the $D$ rest frame; the particle identification 
likelihood ratios of charged tracks in the final state; and, for the $D$ decay modes with a $\pi^0$, the smaller of the two daughter photons' energies.
The cut on the network output variable is optimized for each $D$ decay mode individually by maximizing $S/\sqrt{S+B}$, 
where $S$ ($B$) refers to the signal (background) yield in the signal region that is defined as 
the $\pm3\sigma$ interval around the nominal $D$ meson mass, where $\sigma$ is the decay-mode-dependent invariant-mass 
resolution and ranges from 4 to 12~\mevcc. After optimization, the $D$ purity within the signal region 
increases from 17\% to 42\% while only around 16\% of signal $D$ candidates are lost. 
We use only the $D$ candidates in the signal region in the remainder of the analysis.
More details about the $D$ selection procedure are given in Ref.~\cite{Zupanc:2013byn}. Neutral 
(charged) $D$ mesons are combined with a charged (neutral) pion candidate to form charged $D^{\ast}$ 
candidates. We keep only $D^{\ast}$ candidates in the $\pm3\sigma$ region around the nominal 
value of the mass difference $m(D^{\ast})-m(D)$.

The $D$ and $D^{\ast}$ candidates are combined with a proton or antiproton and a remaining charged pion 
candidates to form $\dppinc$ combinations that represent a sample of inclusively reconstructed charm baryons. 
A kinematic fit to each $\dppinc$ candidate is performed in which the particles are required to originate
from a common point inside the IP region and the $D$ mass is constrained to the nominal value~\cite{Beringer:1900zz}.
We divide the reconstructed charm baryons into the right sign (RS) $D^{(\ast)-}\overline{p}\pi^+$ 
and wrong sign (WS) $D^{(\ast)-}p\pi^-$ and $D^{(\ast)+}\overline{p}\pi^-$ charge combinations 
based on the charm quantum number and baryon number of the $\dppinc$ combinations relative to their total 
electric charge. 
The WS sample, by definition, cannot contain correctly 
reconstructed $\lcp$ candidates so it is used to study properties of the background. 
We retain inclusively reconstructed $\lcp$ candidates with $2.0$~\gevcc~$<\mmiss(\dppinc)<2.5$~\gevcc.
In 15\% of the events, we find more than one $\dppinc$ candidate; in such cases we select at random a single RS (WS) 
candidate for further analysis,
if only RS (WS) candidates are found, or a single RS and a single WS candidate, if RS and 
WS candidates are found in an event.

\begin{figure}[t]
 \centering
 \includegraphics[width=0.49\textwidth]{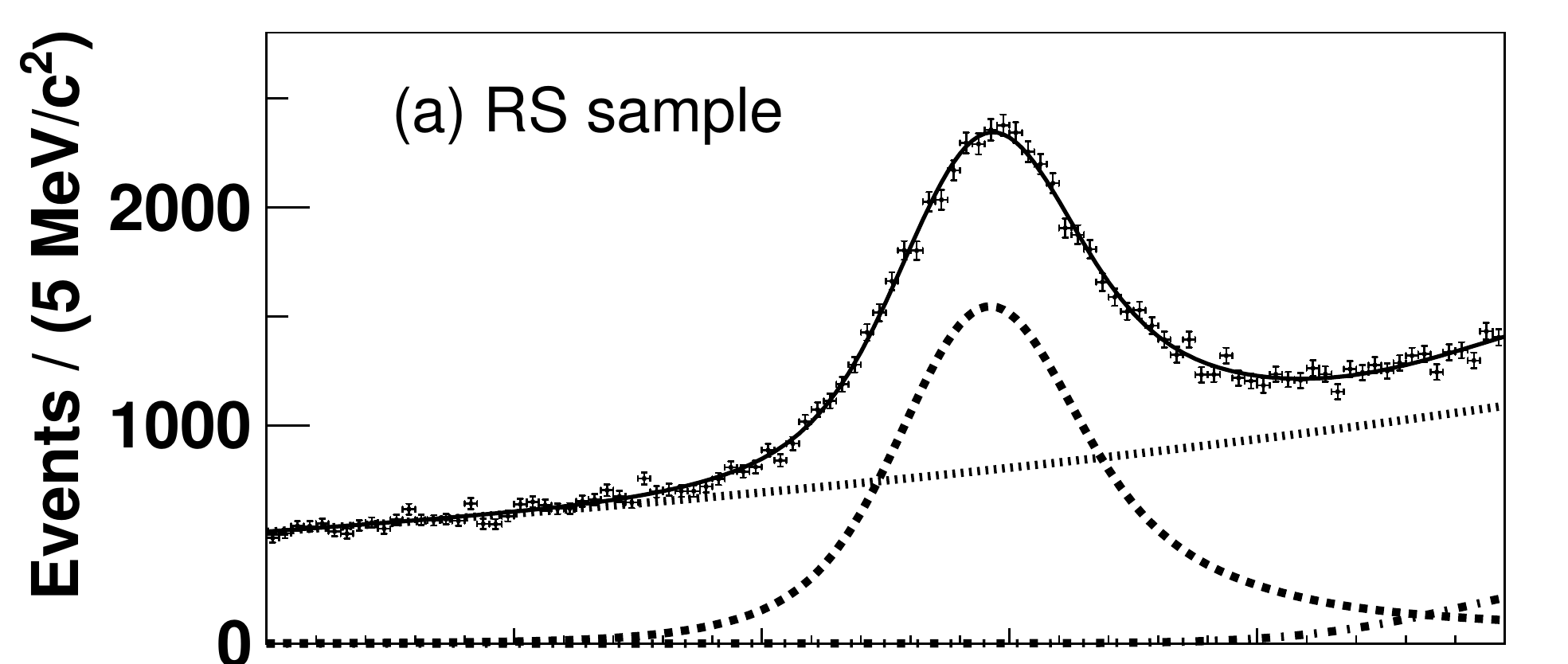}\\
 \includegraphics[width=0.49\textwidth]{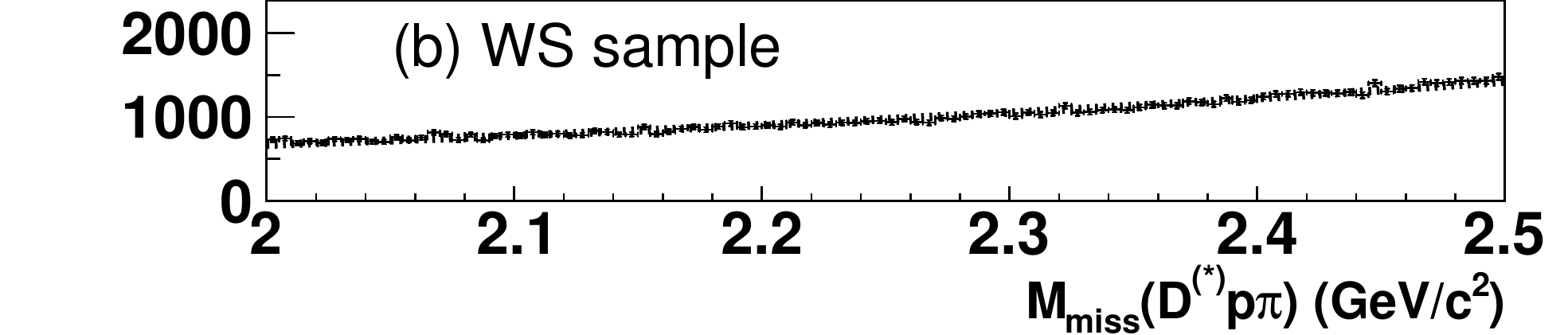}
 \caption{The $\mmiss(\dppinc)$ data distributions (points with error bars) for inclusively reconstructed $\lcp$ baryons from the (a) RS and (b) WS samples with superimposed
 fit results (solid line). The contributions of signal, combinatorial and missing $X$ background are shown with the dashed, dotted, and
 dashed-dotted lines, respectively.}
 \label{fig:inclusive_lc}
\end{figure}
Figure~\ref{fig:inclusive_lc} shows the distributions of $\mmiss(\dppinc)$ for 
RS and WS candidates. A prominent peak at the nominal $\lcp$ mass is visible in 
the spectrum of the RS sample, while the spectrum of the WS sample
is featureless. The yield of inclusively reconstructed $\lcp$ 
baryons is determined by performing a binned maximum likelihood fit to the 
$\mmiss(\dppinc)$ distribution of RS candidates. The inclusively reconstructed 
$\lcp$ candidates fall into three categories: correctly 
reconstructed $\dppinc$ combinations from signal events (signal); correctly reconstructed 
$\dppinc$ candidates from $e^+e^-\to \dppilc X$ events, where $X$ represents one or two 
additional particles produced in the process of hadronization that are missed in the 
reconstruction (missing $X$ background); and all other combinations (combinatorial 
background), which also contribute to the WS sample.

The signal candidates are parameterized as the sum of two components a core and an 
upper-tail part to describe the contribution of events with an undetected initial state 
radiation (ISR) photon~\cite{Benayoun:1999hm}.
The core (upper-tail) component of the signal is described with the sum of 
two (one) Gaussian functions (function) and a bifurcated Gaussian function. 
In the fit, we fix all parameters, including the fraction of ISR events, 
to the values determined from the MC sample except for the means and 
the common resolution scaling factor of the first and the second Gaussian 
function.
The missing $X$ background is parameterized as 
the sum of two Gaussian functions, the first for the case of one missing particle, and the second 
for the case of two missing particles. All the fit parameters except the normalization 
are fixed.
We use an exponential function to describe the combinatorial background, where the single shape 
parameter is fixed to the value determined by the fit to the $\mmiss(\dppinc)$ distribution in 
the WS sample~\footnote{The MC shows that the WS sample correctly models the combinatorial 
background in the RS sample.}. The results of the fits for the WS and RS samples are shown in 
Fig.~\ref{fig:inclusive_lc}. The number of inclusively reconstructed $\lcp$ baryons is found to 
be $N_{\textrm{incl}}^{\Lambda_c} = 36447\pm432$, where the uncertainty is statistical only.

\begin{figure}[t]
  \includegraphics[width=0.9\textwidth]{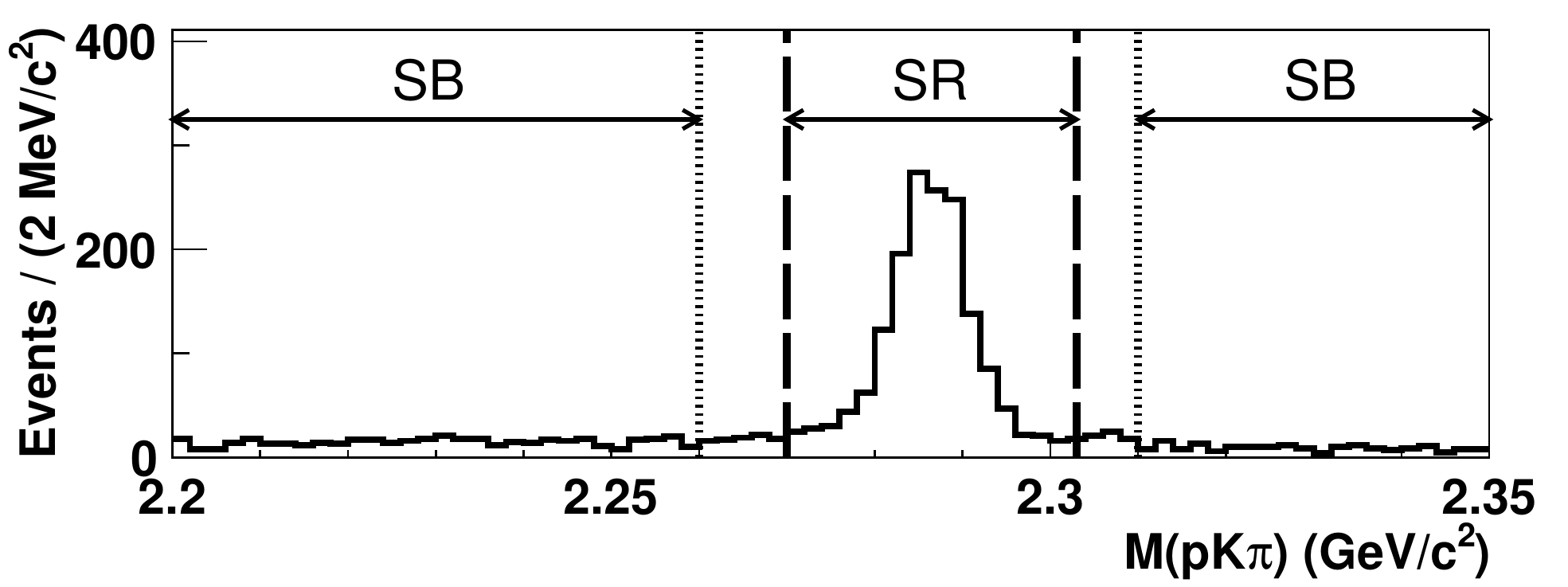}
 \caption{The $M(pK\pi)$ distribution of exclusively reconstructed $\lcpkpi$ candidates within the 
 inclusive $\lcp$ sample. The dashed (dotted) vertical lines indicate the borders of signal (SR) 
 and sideband (SB) regions.}
 \label{fig:exclusive_lc_mass}
\end{figure}

After reconstructing the inclusive sample of $\lcp$ baryons, we proceed with the reconstruction 
of $\lcpkpi$ decays within the inclusive $\lcp$ sample. This is performed by requiring exactly three 
charged tracks to be present in the rest of the event with a net total charge equal to the 
charge of the inclusively reconstructed $\lcp$ candidate. The track whose charge is opposite that of 
the inclusive $\lcp$ candidate is assigned to be the kaon. From the two same-sign tracks, we 
identify the proton based on the PID likelihood ratios; 
the remaining track is assumed to be a pion. Figure~\ref{fig:exclusive_lc_mass} shows the 
invariant-mass distribution of exclusively reconstructed $\lcpkpi$ decays within the inclusive 
$\lcp$ sample. A clear peak at the nominal mass of the $\lcp$ can be seen above a very low 
background. 

MC studies show that the $\lcp$ inclusive reconstruction efficiency depends weakly on the $\lcp$ decay mode 
and therefore the inclusively reconstructed $\lcp$ sample does not represent a truly inclusive sample 
of $\lcp$ baryons. This effect is described with the factor 
$\fbias = {\varepsilon^{\rm inc}_{\lcp\to f}}/{\overline{\varepsilon}{}_{\lcp}^{\rm inc}}$ in Eq.~(\ref{eq:abs_br})
and is given by the ratio of $\lcp$ inclusive reconstruction efficiency for $\lcp\to f$ decays,
$\varepsilon^{\rm inc}_{\lcp\to f}$, and the average $\lcp$ inclusive reconstruction efficiency, 
$\overline{\varepsilon}{}_{\lc}^{\rm inc}=\sum_i {\cal B}(\lcp\to i)\varepsilon^{\rm inc}_{\lc\to i}$. 
In the case of $f=\pkpi$, the $\fbias$ value determined by MC that includes all known $\lcp$ decays 
is found to be consistent with unity and the product of the tag bias 
and the exclusive reconstruction efficiency is 
$\fbias\varepsilon(\lcpkpi)=(54.5 \pm 0.6)\%$, 
where the uncertainty is due to the limited MC statistics.

\begin{figure}[t]
\includegraphics[width=0.485\textwidth]{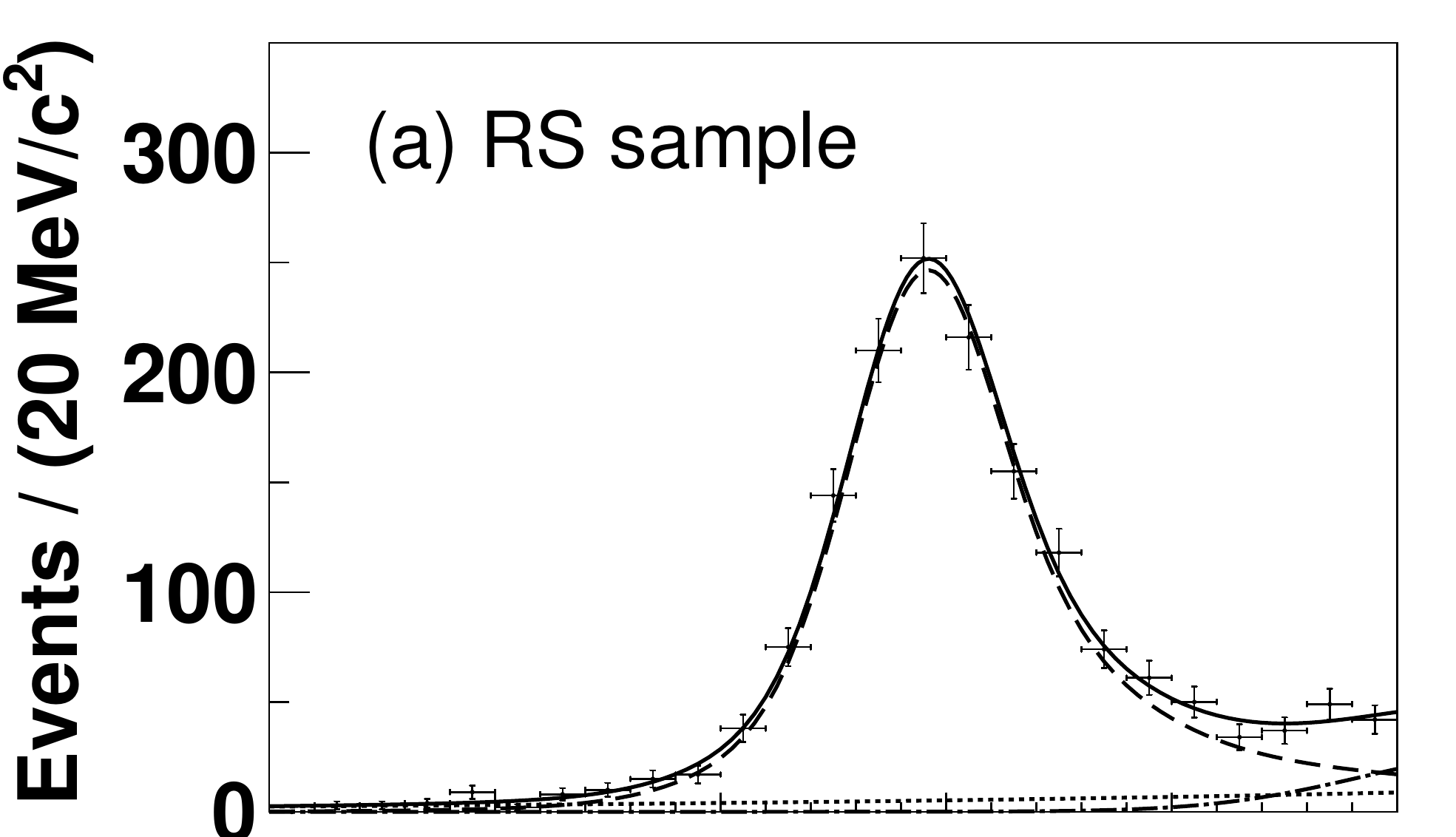}
\includegraphics[width=0.485\textwidth]{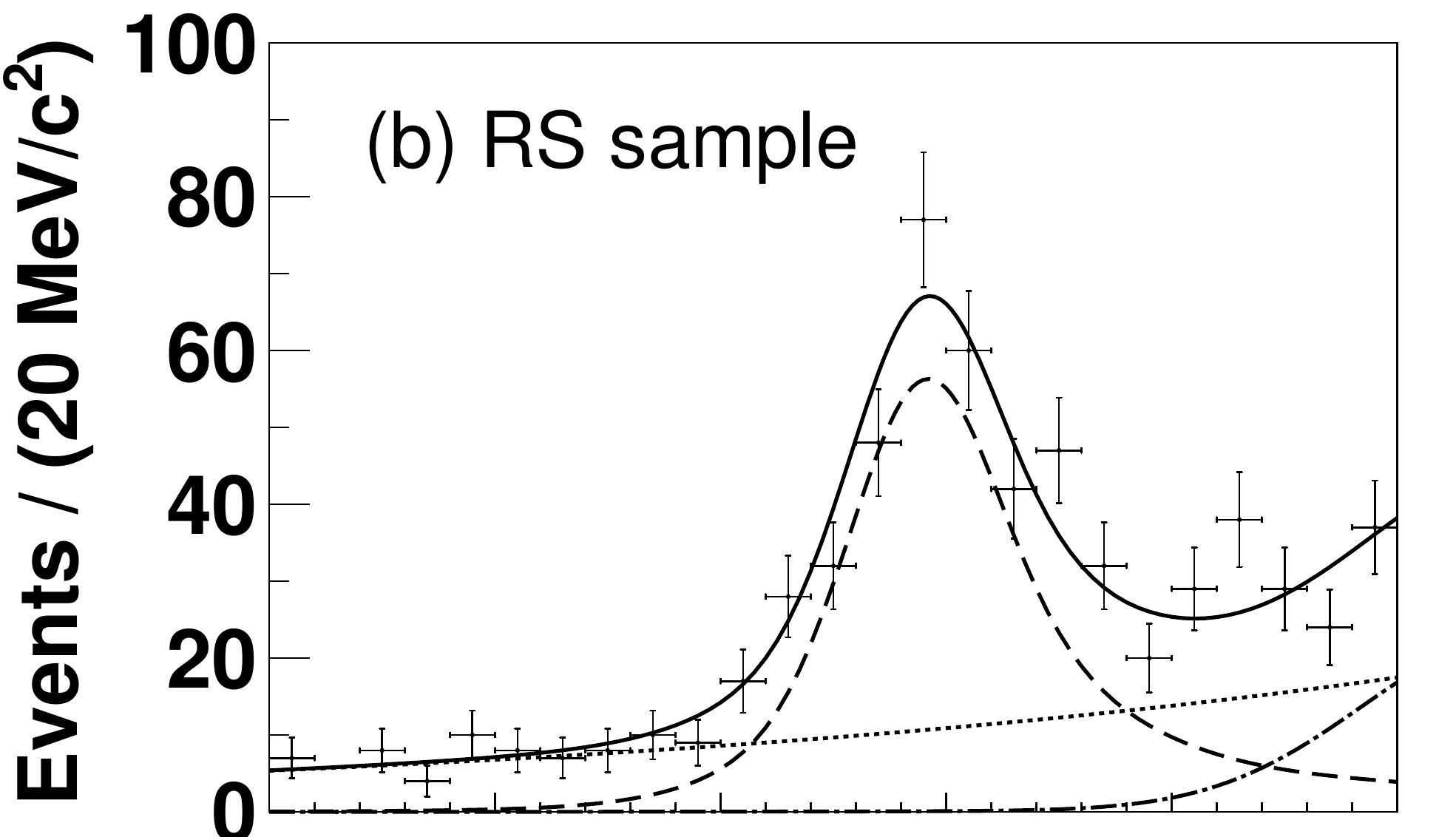}\\
\includegraphics[width=0.485\textwidth]{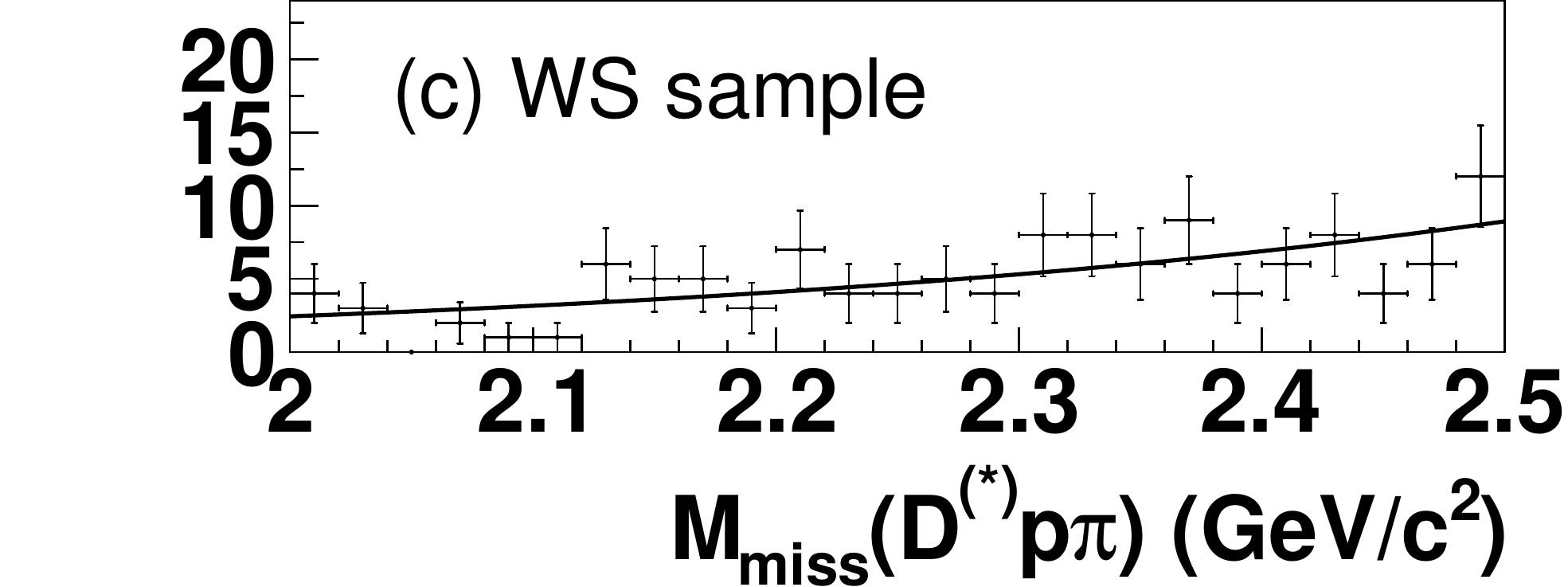}
\includegraphics[width=0.485\textwidth]{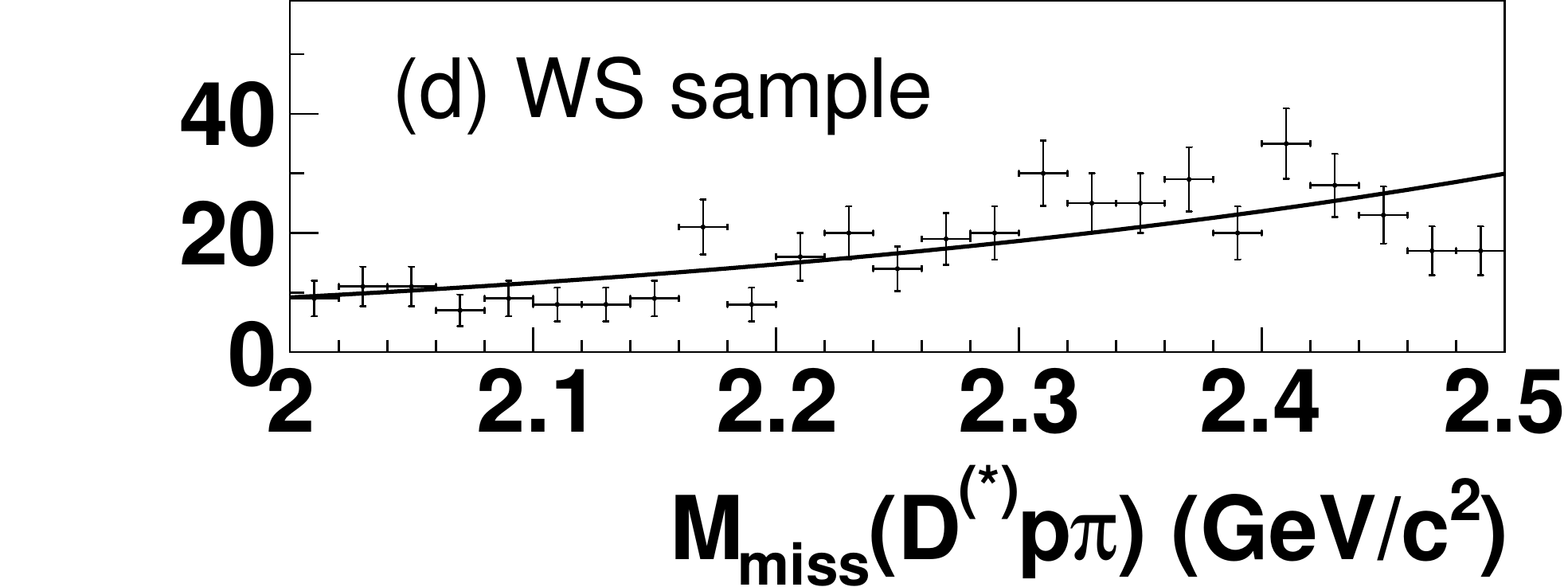}\\
 \caption{The $\mmiss(\dppinc)$ data distributions (points with error bars) of exclusively reconstructed $\lcpkpi$ candidates. (a) and (c) for the SR region and (b) and 
 (d) SB region of $M(pK\pi)$ for the RS and WS samples, respectively, with superimposed fit results (solid line). The contributions of signal, 
 combinatorial and missing $X$ background are shown with the dashed, dotted, and dashed-dotted lines, respectively.}
 \label{fig:lc_exclusive}
\end{figure}

The number of exclusively reconstructed $\lcpkpi$ decays within the inclusive $\lcp$ sample is 
determined by performing a fit to the $\mmiss(\dppinc)$ distribution of candidates within the 
signal (SR) and sideband (SB) regions of $M(pK\pi)$. The main reason to perform a fit to the
$\mmiss(\dppinc)$ distribution rather than the $M(pK\pi)$ distribution is that, in the former 
case, the systematic uncertainty related to the parameterization of $\mmiss(\dppinc)$ distributions 
cancels to a large extent in the ratio of exclusively and inclusively reconstructed $\lcp$ 
candidates (see Eq.~\ref{eq:abs_br}) while, in the latter case, it does not. The fits to the 
$\mmiss(\dppinc)$ distributions of candidates within the signal and sideband regions of $M(pK\pi)$ 
are performed in the same way and using the same parameterization as the fit to the $\mmiss(\dppinc)$ 
distribution of all inclusive $\lcp$ candidates. We first fit candidates in the WS sample to 
determine the shape parameter of the combinatorial background that we then fix in the fit of 
the RS sample. In the RS fit, the signal shape parameters are fixed to the values found in the 
fit to the total inclusive $\lcp$ sample. Figure~\ref{fig:lc_exclusive} shows the results of the fits to the RS and WS 
$\mmiss(\dppinc)$ distributions of exclusively reconstructed $\lcp$ candidates within the 
signal and sideband regions of $M(pK\pi)$. 
The signal yields are found to be 
$N_{\textrm{excl}}^{\textrm{SR}} = 1457\pm44$ and $N_{\textrm{excl}}^{\textrm{SB}} = 332\pm27$, 
where the uncertainties are statistical.

The number of exclusively reconstructed $\lcpkpi$ decays within the inclusive $\lcp$ sample 
(where both exclusive and inclusive $\lcp$ candidates are correctly reconstructed) is given 
by $N(\lcpkpi)=N_{\textrm{excl}}^{\textrm{SR}}-r^{\textrm{SB}}_{\textrm{SR}}N_{\textrm{excl}}^{\textrm{SB}}$,
where $N_{\textrm{excl}}^{\textrm{SR(SB)}}$ is the yield of correctly reconstructed inclusive 
$\lcp$ candidates from the fit to the $\mmiss(\dppinc)$ distribution of candidates within the
signal (sideband) region of the $M(pK\pi)$ distribution. The ratio $r^{\textrm{SB}}_{\textrm{SR}}$ 
is formed from the yields of correctly reconstructed inclusive $\lcp$ candidates but wrongly reconstructed 
exclusive $\lcp$ candidates within the signal and sideband regions. These candidates peak in 
$\mmiss(\dppinc)$ but not in $M(pK\pi)$. The ratio is determined on a 
simulated sample of events to be $r^{\textrm{SB}}_{\textrm{SR}}=0.296 \pm 0.015$. The number of exclusively reconstructed 
$\lcpkpi$ decays is thus $N(\lcpkpi)=1359\pm45$, where the uncertainty includes the  
$N_{\textrm{excl}}^{\textrm{SR}}$ and $N_{\textrm{excl}}^{\textrm{SB}}$ statistical uncertainties.
The branching fraction, given by Eq.~\ref{eq:abs_br}, is $\br(\lcpkpi)=(6.84\pm0.24{})\%$, 
where the uncertainty includes both exclusive ($N(\lcpkpi)$) and inclusive ($N_{\textrm{incl}}^{\Lambda_c}$) uncertainties.

As a check, we extract the branching fractions of $\lcpkpi$ decays for each $D^{(\ast)+}$ decay mode individually; these are found
to be in good agreement with each other as well as with the nominal result. 
As mentioned above, the alternative way to determine the number of exclusively reconstructed 
$\lcpkpi$ decays is to perform a fit to the $M(pK\pi)$ distribution: we find $1208\pm41$ correctly reconstructed
$\lcpkpi$ decays within the $2.16$~\gevcc~$<\mmiss(\dppinc)<2.38$~\gevcc region and the resulting
branching fraction, $(6.78\pm0.24{})\%$, in excellent agreement with the nominal result.
We perform another model-independent measurement of $\br(\lcpkpi)$ in events of 
$e^+e^-\to D^{(\ast)-}\overline{p}\pi^+\lcp$ and $e^+e^-\to \overline{D}{}^{0}\overline{p}\lcp$ 
that utilizes a cut-based $D^{(\ast)}$ selection. Here, we determine the number of 
$\lcpkpi$ decays within the inclusive $\lcp$ sample by a fit to the missing 
energy of the event, which is expected to be zero for signal. 
The measured branching fraction, $(7.04\pm0.38)\%$, where the uncertainty is statistical, is 
found to be in good agreement with the nominal result.

\begin{table}[t!]
\centering
\caption{Summary of systematic uncertainties.}
\label{tab:systematics}
\begin{ruledtabular}
\begin{tabular}{lc} 
Source 		& Uncertainty [\%] \\ \hline
Tracking 		& 1.1\\
Proton ID		& 0.4\\
Efficiency		& 1.1\\
Dalitz model            & 1.1\\
$f_{\textrm{bias}}$      & 1.5\\
Bkg. subtraction 	& ${}^{+0.5}_{-0.9}$\\ 
Fit Model		& ${}^{+1.7}_{-2.9}$\\ \hline
Total			& ${}^{+3.0}_{-3.9}$\\ 
\end{tabular}  
\end{ruledtabular}
\end{table}
Systematic uncertainties in the calculation of the branching fraction arise due to imperfect knowledge of the 
efficiency of the exclusive reconstruction of $\lcpkpi$ decays within the inclusive $\lcp$ sample and the modeling 
of signal and background contributions in fits to the $\mmiss(\dppinc)$ distributions of inclusive and exclusive candidates. 
Since the branching fraction is determined relative to the number of inclusively reconstructed $\lcp$ baryons, the systematic
uncertainties in the reconstruction of the $\dppinc$ system cancel. 
The estimated  systematic uncertainties are summarized in Table~\ref{tab:systematics} and described below. 
The systematic uncertainty due to charged-track reconstruction efficiency
is estimated to be 0.35\% per track (1.1\% in total) from partially reconstructed $D^{\ast +} \to D^0(K^0_S\pi^+\pi^-)\pi^+$
decays. We estimate the uncertainty due to proton identification (0.4\%) using a $\Lambda\to p\pi^-$ sample. 
The systematic uncertainty due to the requirement that there are no additional tracks present in an event
after the exclusive reconstruction of $\lcpkpi$ candidates within the inclusive $\lcp$ sample is estimated as follows.
We compare the ratios of events with correctly reconstructed exclusive $\lcp$ candidate within the inclusive 
$\lcp$ sample with and without any additional tracks detected as determined on simulated and data samples. 
We find the ratios to be small and in good agreement and therefore assign no additional systematic uncertainty.
We include, as 
a source of systematic uncertainty, the statistical uncertainty of the MC-determined efficiency (1.1\%). The reconstruction 
efficiency of $\lcpkpi$ decays is found to vary weakly across the $pK^-\pi^+$ Dalitz distribution. In calculating $\br(\lcpkpi)$, 
we use the Dalitz-plot-integrated MC efficiency. The decay amplitude in the MC is the incoherent sum of all known resonant two-body 
contributions. We vary the relative contributions of these intermediate states within their uncertainties~\cite{Beringer:1900zz} to
estimate the systematic uncertainty due to the Dalitz model to be 1.1\%. 
Possible differences in relative rates of individual $\lcp$ decay modes between MC simulation 
and data that impact the $\fbias\varepsilon(\lcpkpi)$ determination are estimated by studying 
the distributions of the number of charged particles 
and neutral pions produced in $\lcp$ decays in MC and data~\cite{Zupanc:2013byn}; the corresponding systematic uncertainty is
estimated to be 1.5\%. 
We propagate the statistical uncertainty 
of the $r^{\textrm{SB}}_{\textrm{SR}}$ ratio and perform the background subtraction using the upper and lower
$M(Kp\pi)$ sidebands only and take the difference from the nominal value 
to estimate the systematic uncertainty due to background subtraction (in total ${}^{+0.5}_{-0.9}\%$).
We estimate the systematic uncertainty due to the $\mmiss(\dppinc)$ fit model by 
varying the shape parameter of the combinatorial background within its uncertainties 
(as obtained from the WS sample fit) $(\pm0.7\%)$; 
using a second-order polynomial to describe combinatorial background 
instead of the exponential function $({}^{+1.5}_{-2.8}\%)$;  
using a parameterization for the one- or two-missing-particle backgrounds 
separately instead of the nominal mixture of the two $(\pm0.07\%)$;
giving an additional contribution to the total fit function that describes 
a possible peaking contribution from 
$e^+e^-\to D^{(\ast)-}\overline{p}\pi^+\Sigma_c(2455/2520)$ events ($\pm0.01\%$); 
varying the signal shape parameters obtained from the fit to the inclusive
sample in fits to the signal and sideband regions of the $M(pK\pi)$ distribution $(\pm0.3\%)$;
and varying the fraction of ISR within the signal model by $\pm20\%$, which is the precision
of the prediction given in Ref.~\cite{Benayoun:1999hm} $(\pm0.3\%)$. 
The total systematic uncertainty is the sum of the above contributions in quadrature.

In summary, we perform the first model-independent measurement of 
the absolute branching fraction of the decay $\lcpkpi$ using 
the Belle final data sample
corresponding to 978~\fb. We measure $\br(\lcpkpi)=(6.84\pm0.24({\textrm{stat.}}){}^{+0.21}_{-0.27}({\textrm{syst.}}))\%$,
which represents a fivefold improvement in precision over previous model-dependent determinations.
This measurement will also improve significantly the precision of the branching fraction of other 
$\lcp$ decays and of decays of $b$-flavored mesons and baryons involving $\lcp$.

\acknowledgments
We thank the KEKB group for excellent operation of the
accelerator; the KEK cryogenics group for efficient solenoid
operations; and the KEK computer group, the NII, and 
PNNL/EMSL for valuable computing and SINET4 network support.  
We acknowledge support from MEXT, JSPS and Nagoya's TLPRC (Japan);
ARC and DIISR (Australia); FWF (Austria); NSFC (China); MSMT (Czechia);
CZF, DFG, and VS (Germany); DST (India); INFN (Italy); 
MOE, MSIP, NRF, GSDC of KISTI, BK21Plus, and WCU (Korea);
MNiSW and NCN (Poland); MES and RFAAE (Russia); ARRS (Slovenia);
IKERBASQUE and UPV/EHU (Spain); 
SNSF (Switzerland); NSC and MOE (Taiwan); and DOE and NSF (USA).

\bibliography{br-LC2pKpi}

\end{document}